\documentstyle[11pt,paspconf]{article}

\begin{document}

\title{Microwave emission by dust: mechanisms, properties and prospects
for ISM studies}

\author{A. Lazarian \altaffilmark{1}}
\affil{Princeton University Observatory, Princeton, NJ 08544}


\altaffiltext{1}{Present address: CITA, Univ. of Toronto, e-mail:
lazarian@cita.utoronto.ca} 



\begin{abstract}
I review my work with Bruce Draine on 
dust emissivity at microwave frequencies (3~cm - 3~mm).
This emissivity explains the recently detected "anomalous" component
of the galactic foreground emission. Both small ($a<10^{-7}$~cm) and
large grains contribute to this emission. Small grains have electric
dipole moments and emit while they rotate; the microwave emission
of large grains is mostly due to magneto-dipole radiation. 
Most efficient  magneto-dipole emitters are strongly magnetic, 
e.g. ferrimagnetic or ferromagnetic, materials. 
The relative role of the two mechanisms can be established through
observations of microwave emissivity from dark clouds. 
New microwave window is a window of
opportunity for interstellar studies. Magnetic fields inside
dark clouds  may be successfully
studied via microwave polarization. Microwave emissivity constrains
the abundance of strongly magnetic materials. For instance,
the available data at 90~GHz indicate that not more than $\sim 5\%$ of
interstellar Fe is in the form of metallic iron grains or inclusions
(e.g., in ``GEMS''). Future missions, e.g. MAP and PLANCK, will bring
a wealth of microwave data that can be successfully used to study
ISM. Such a study would be appreciated by cosmologists who franticly
try to remove all foregrounds from their data.
\end{abstract}




\section{Introduction}

Experiments to map cosmic microwave background radiation have
stimulated recent studies of diffuse Galactic emission. 
Kogut et al. (1996), Oliveira-Costa et al. (1997) and Leitch 
et al. (1997) have reported a new component of galactic microwave
emission which is correlated with the thermal emission from dust.

This emission cannot originate from plasma with $T<10^4$~K, as 
this would result in 60 times the observed amount of H$_{\alpha}$
(Leitch et al. 1997). If, in order to escape the H$_{\alpha}$
constrain, one assumes that the plasma temperature is greater
than $10^{6}$~K, one faces impossible energy requirements.
Indeed, the required energy injection rate is at least 2 orders of magnitude
greater than is expected from supernova explosions (Draine \& Lazarian 1998a).

As this radiation is correlated with the dust 100~$\mu$m thermal emission,
it is natural to assume it to be related to dust. The
study by Draine \& Lazarian (1998a,b; 1999) have revealed two new mechanisms
of dust microwave emission: one based on electric dipole
emission from small rotating grains, the other on the magneto-dipole
emission.

In what follows we discuss rotational emission from small grains (section~2),
magneto-dipole emission from large grains (section~3), analyse
the observational ways to discriminate between the two mechanisms (section~4),
discuss how microwave studies constrain grain properties (section~5) and
outline perspectives of the microwave studies of dust (section~6).
The discussion is intentionally made qualitative. The corresponding
quantitative results are published in Draine \& Lazarian (1998 a,b; 1999).

\section{Small rotating grains}

A grain with a dipole moment $\mu$ rotating at an angular velocity 
$\omega$ radiates power:
\begin{equation}
P=\frac{2}{3}\frac{\omega^4 \mu^2 \sin^2\theta}{c^3}
\end{equation}
where $\theta$ is the angle between the angular velocity and dipole
moment vectors. One source of the grain dipole moment is grain electric
charging
as,
in general, the centroids of charge and mass are displaced. However,
it is shown in Draine \& Lazarian (1998b) that intrinsic dipole moment
arising from the polarization of chemical bonds is more important. 
For the purpose of modeling, it seems possible to assume that 
grain material has short-range order and consists of randomly-arranged
chemical substractures. In this case $\mu\approx N^{1/2}\beta$, where
$N$ is the number of substractures and the magnitude $\beta\approx 0.4$
can be used as a representative value.

The rotational velocity $\omega$ is determined by grain interactions
with ions, neutrals, and with electromagnetic radiation. Calculations
in Draine \& Lazarian (1998b) have shown that interactions with ions
are more important for grain dynamics than grain interactions with neutrals
even when ion density is much less than that of neutrals. 
This is the consequence of the fact that 
ions can be attracted to the grain surface and this focusing dramatically
enhances both the grain effective cross section and the value of
angular momentum that is being transfered per collision. Moreover,
grains also interact with passing ions via grain electric dipole 
moment. Such ``distant'' collisions provide efficient coupling of grains
to plasma.
The grain interaction with radiation serves mostly as a source of grain
damping, although emission of infrared photons can also excite
grain rotation. The dipole rotational emission at microwave frequencies
results in non-linear damping of grain rotation.

The observed microwave emissivity is proportional to the number of emitting
grains along the line of sight. MRN distribution (Mathis, Rumpl
\& Nordsieck 1977), extended down to very small sizes, might provide 
an estimate for the number of very small grains. In fact, it
is known (Leger \& Puget 1984, Draine \& Anderson 1985) that the MRN
underestimates the number density of very small grains. In  Draine \&
Lazarian (1998b) model  assumes that 5\% of interstellar
carbon is locked in grains with $a<10^{-7}$~cm, which is rather conservative
estimate (compare Boulanger \& Perault 1988).

It is shown in Draine \& Lazarian (1998a) that small grains can 
reproduce well the ``anomalous'' foreground emissivity discovered
by observations. The advantage of this explanation is that it appeals
to small grains known to be present in the ISM and that it
uses
 conservative assumptions about their properties.

\section{Magnetic Grains}

An alternative explanation for the observed microwave emission is
based on magneto-dipole emissivity (Draine \& Lazarian 1999). 
The corresponding grains should be large, as the intensity of the
emission is proportional to the mass of grain material and most
of dust mass is in the form of large grains. They also should be
strongly magnetic, as only strongly magnetic grain are efficient
microwave emitters.

First of all, it is necessary to clarify why magneto-dipole emission, 
 marginal at optical and infrared frequencies ($\nu>10^{12}$~Hz), 
is important at the microwave bandwidth. Most materials have negligible
magnetic response to oscillating magnetic fields at frequencies
$>10^{11}$~Hz, but when the frequency of the field
approaches the precession frequency of an electron spin in the 
magnetic field of its neighbors, the magnetic response becomes
appreciable. In fact, Draine \& Lazarian (1999) have shown that
in the range of $10-30$~GHz the emission arising from
thermal fluctuations of magnetization for ordinary paramagnetic
dust grains is higher than the emission arising from
electric ``vibrational'' transitions within the dust.

Another way to see that low frequency magnetic response should
dominate at the microwave range is to analyse Kramers-Kroning relations
(see Draine \& Lazarian 1999). Indeed, if the zero-frequency magnetic and
electric susceptibilities of a material
are comparable while its high frequency magnetic
response is negligible, Kramers-Kroning relations suggest that
the magneto-dipole adsorption should dominate at low frequencies.

Observations of the interstellar depletions indicate that
$\sim 10\%$ of atoms in grains are Fe and therefore it would not
be surprising if some fraction of interstellar grain population is 
strongly magnetic (see Jones \& Spitzer 1967). However, candidate
materials e.g. metallic iron/nickel, magnetite (Fe$_3$O$_4$),
and maghemite ($\gamma$Fe$_2$O$_3$) do not provide a reasonable fit
to the data. A hypothetical material giving a good fit should have
 magnetic response weaker than 
metallic iron, but stronger than ferrimagnetic magnetite. It 
could be, for example, Fe/Ni alloy with an appreciable concentration
of O, H, Si and other impurities. At the moment we cannot exclude
that such material indeed makes appreciable contribution to the
observed microwave emission. The relative role of the magneto-dipole
emissivity and emissivity from small rotating grains should be 
determined on the basis of future observations. 

\section{Dense Clouds as a Test Case}

Measurements of polarization can provide one way to distinguish
microwave emission from rotating small grains and magneto-dipole
emission from hypothetical strongly magnetic grains. However,
measurements of intensity from dense clouds may enable an
easier way to discriminate between the two emission mechanisms.

It is well known that the fraction of small grains is reduced
in dense clouds. Therefore the microwave emissivity per hydrogen 
atom from rotating small grains is expected to be reduced as 
compared to the diffuse media. Magneto-dipole emissivity 
depends neither on grain size nor on the abundance of small grains.
As the result,higher if magneto-dipole transitions are
responsible for the ``anomalous'' microwave emissivity the expected 20-90~GHz
intensity from dark clouds is higher compared to the spinning dust case.

Adopting the Lazarian, Goodman \& Myers (1997)
model of L1755 dark cloud, Draine \& Lazarian (1998b)
obtained expected antenna
temperatures of $\sim 1$~mK at 10-30~GHz if the emission is due to
spinning grains. This is the level of sensitivity that is required
to exclude magneto-dipole origin of ``anomalous'' microwave emission.

\section{Constraining Grain Properties}

Microwave emission can provide a valuable tool for dust studies.
First of all, microwave emissivity can trace the variation of 
small grain abundance as cloud density increases. It also can
constrain magnetic properties of dust materials. The latter is
important for solving the mystery of grain alignment. Indeed,
if grains are strongly magnetic, Davis-Greenstein relaxation (1951)
should be sufficient to insure alignment. Even with the limited present
day data it is possible to say that no more than 5\% of interstellar 
Fe is in the form of metallic grains or inclusions (e.g. in ``GEMS'') 
(Draine \& Lazarian 1999). To constrain magnetic properties of
interstellar dust 
materials further, both laboratory studies of candidate materials
and sensitive observations in $10-100$~GHz frequency range are
required.

\section{Studies of Microwave Background}

Studies of CMB are hot area of research at present. The microwave 
foreground arising from dust is a nuisance for cosmology oriented
people who try to study the evolution of 
the early Universe. Their extensive studies,
both ground and space based, should provide a wealth of valuable
microwave data. To filter out dust microwave contribution cosmologists
will have to  understand its  properties better and the
corresponding research can be a gold mine for the ISM studies.
In my other paper in this volume  (Lazarian 1999) I
discuss the origin of microwave polarization and how it can help the 
studies of ISM magnetic fields. 

\section{Summary}

~~~~~1. Grains smaller than $\sim 10^{-7}$~cm rotate at frequencies higher 
than 10~GHz. Small grains are likely to have non-zero electric dipole
moment. As a result, these grains emit at microwave frequencies.
This emission provides the most likely explanation for the 
 detected recently ``anomalous'' galactic foreground.

2. For typical paramagnetic interstellar grains, characteristic 
frequency of magneto-dipole transitions lies in the microwave range.
Therefore magneto-dipole emissivity from these grains dominates their
vibrational electric dipole emissivity at $\sim 10-30$~GHz.
If grains are strongly magnetic, they may provide an alternative
explanation for the  ``anomalous'' galactic foreground.

3. Magneto-dipole emissivity is proportional to dust mass. 
Emissivity from rotating grains lies in the microwave range
only for ultra small grains . 
Therefore this emissivity per hydrogen atom should
be lower in dark clouds, where the abundance of small grains is
reduced. This provides a way to measure the relative importance
of the two candidate mechanisms of microwave emission.

4. Intensive ongoing research of microwave background can provide
 valuable information both on dust properties and on the ISM magnetic
field structure.

\acknowledgments
The present qualitative discussion is based on the quantitative 
work done with Bruce Draine. I acknowledge discussions with 
Dick Crutcher, Chris McKee and Phil Myers. The research was supported
by NASA grants NAG5-2858, NAG5-7030, and CITA Senior Fellowship.


%
%

%


\begin{references}
\reference Boulanger, F., \& Perault, M. 1988, ApJ, 330, 964
\reference de Oliveira-Costa, A., Kogut, A., Devlin, M.J., Netterfield, C.B.,
Page, L.A., \& Wollack, E.J. 1997, ApJ, 482, L17
\reference Davis, L., \& Greenstein, J. 1951, ApJ, 114, 206
\reference Draine, B.T., \& Lazarian, A. 1998a, ApJL, 494, L19
\reference Draine, B.T. \& Lazarian, A. 1998b, ApJ,  508, 000
\reference Draine, B.T., \& Lazarian, A. 1999, ApJ, in press (also
astro-ph/9807009)
\reference Jones, R.V., \& Spitzer, L. 1967, ApJ, 147, 943
\reference Kogut, A., Banday, A.J., Bennett, C.L., Corski, K.M., 
Hinshaw, G. \& Reach, W.T. 1996, ApJ, 460, 1
\reference Lazarian, A. 1999, contribution to this volume
\reference Lazarian, A., Goodman, A.A., \& Myers, P.C. 1997, ApJ, 490, 273
\reference Leger, A., \& Puget, J.L. 1984, ApJ, 278, L19
\reference Leitch, E.M., Readhead, A.C.S., Pearson, T.J., \& Myers, S.T.
1997, ApJ, L23
\reference Mathis, J.S., Rumpl,W., \& Nordsieck, K.H. 1977, ApJ, 217, 425
\end{references}
\end{document}